\documentclass[sort&compress,
    ,final            
  ]
  {aipproc}
\layoutstyle{8x11double}
\usepackage{psfrag}   
\usepackage{amsmath} 
\usepackage{amsbsy} 
\newcommand{\ww}[1]{\underline{\underline{{\bf #1}}}}
\newcommand{\bma}{\begin{bmatrix}}
\newcommand{\ema}{\end{bmatrix}}
\newcommand{\be}{\begin{equation}}
\newcommand{\ee}{\end{equation}}
\begin{document}

\title{
Incremental response of granular materials: DEM results
}


\author{F. Froiio}{
  address={Ecole Centrale de Lyon, LTDS, 36 av Guy de Collongue, 69134
  Ecully Cedex, France}
}

\author{J.-N. Roux}{
  address={Universit\'e Paris-Est, U.R. Navier, LMSGC, 2 all\'ee Kepler,
  77420 Champs-sur-Marne, France}
}
\begin{abstract}
We systematically investigate the incremental response of various
equilibrium states of dense 2D model granular materials, along the
biaxial compression path ($\sigma_{11}<\sigma_{22}$,
$\sigma_{12}=0$).  Stress increments are applied in arbitrary directions
in 3-dimensional stress space $(\sigma_{11},
\sigma_{22},\sigma_{12})$. In states with stable contact networks
we compute the stiffness matrix and the elastic moduli, and separate
elastic and irreversible strains in the range in which the
latter are homogeneous functions of degree one of stress increments.
Without principal stress axis rotation, the response abides by
elastoplasticity with a Mohr-Coulomb criterion and a non-associated flow
rule. However a nonelastic shear strain is also observed for increments of $\sigma_{12}$,
and shear and in-plane responses couple.
This behavior correlates to the distribution of friction
mobilization and sliding at contacts. 
\end{abstract}

\maketitle


\section{Introduction}
Although the mechanical behavior of solidlike granular materials under quasistatic loading conditions is often modeled
as elastoplastic at the continuum level~\cite{VER98,MIT93}, there are still few studies addressing the microscopic origins of
such a behavior by discrete, grain-level simulation~\cite{AMLHV05,TCV05,DSDN07,R08}. 
To assess the applicability of elastoplastic laws,
one needs to investigate the response to small stress or strain increments, superimposed in various directions on an equilibrium state.  
One essential motivation for such studies is the prediction of shear band formation, for which such criteria as the
Rudnicki-Rice~\cite{VaSu95} condition involve the incremental response. In particular, localization is crucially sensitive to the response to
stress increments with rotation of principal axes, as when some simple shear is superimposed on a biaxial compression~\cite{DeCh02}.
The present study addresses this issue for the simplest model material, an assembly of disks in 2 dimensions, for which the
response to load increments in all 3 dimensions of stress space is computed at various points along a biaxial loading path.
\section{Model material and methods}
Our simulation samples comprise 5600 disks enclosed in a periodic rectangular cell. The diameter distribution is uniform between
$0.7d$ and $1.3d$. We use a simple, frictional-elastic contact model, involving (constant) normal contact stiffness $K_N$, tangential contact
stiffness $K_T$ (here we set $K_T=K_N$) and a friction coefficient, $\mu$, set to $0.3$. The normal (elastic) contact force is 
$F_N = K_N\, h$  where $h$ is the interpenetration of contacting disks (which models surface deflection). 
The tangential force $F_T$ relates to the elastic part $\delta$ of the tangential relative displacement, as $F_T=K_T\delta$,
and is incrementally computed to enforce the Coulomb condition $|F_T| \le \mu F_N$. Some viscous damping is also introduced,
which proves irrelevant to the material behavior for low enough strain rates.

We focus here on dense samples, which are initially assembled without friction, under an isotropic pressure $P$. The initial state 
is thus characterized by an isotropic fabric and a large coordination number (close to 4). 
The dimensionless stiffness parameter $\kappa = K_N/P$ sets the scale of contact deflections, as $h/d\propto \kappa^{-1}$. We choose
value $\kappa=10^4$ in most simulations. 

Deformations of the simulation cell, i.e. macroscopic strains, are controlled, or vary in response to applied stresses. This is achieved with specific
implementations of Parrinello-Rahman and Lees-Edwards techniques (first developed for molecular systems~\cite{AT87}), as explained in Ref.~\cite{PR08a}. 
Stresses are given by the classical Love formula.
In the biaxial compression test, the deformable cell remains rectangular, its edges parallel to the principal stress
directions. Principal stress value $\sigma_1$ (the lateral stress) is kept equal to $P$, while $\sigma_2$ (the axial stress)
increases in response to strain $\epsilon_2$, which grows at a controlled rate 
(compressive stresses and shrinking strains are positive). As
indicated in Fig.~\ref{fig:biax}, the compression test is stopped at different stages
and the sample is equilibrated at constant stresses. This entails slight \emph{creep} strain increments,
which remain quite small (of order $10^{-6}$), until equilibrium conditions are satisfied with good accuracy (the tolerance is $10^{-4}$ in units of
$P$, $dP$, and $d^2P$ for stresses, forces and moments, respectively). 
In those well-equilibrated intermediate states, hereafter referred to as \emph{investigation points},
we first compute elastic moduli. To do so, we use
the stiffness matrix associated to the contact network, as in Ref.~\cite{iviso3}. 
It is convenient to denote stresses and strains as 3-vectors (as $\delta\vec\sigma$, $\delta\vec\epsilon$)
with $\delta \sigma_3 = \sqrt{2} \sigma_{12}$, while notations $\delta\sigma_1$,
$\delta\sigma_2$ keep the same meaning (and similarly for $\delta\vec\epsilon$). 
Due to symmetry about the principal axes there are four independent elastic moduli, 
which satisfy:
\begin{figure} 
\includegraphics[angle=270,width=6cm]{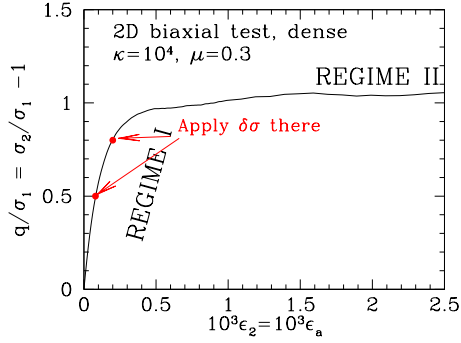}
\caption{\label{fig:biax} Deviator stress versus axial strain curve, and 
location of 2 investigation points for incremental response} 
\end{figure}
\be
\delta\vec\sigma = \ww{C}\cdot\delta\vec\epsilon^{E}, \mbox{ with } \ww{C}=
\bma C_{11} & C_{12} &0\\ C_{12} & C_{22} &0\\ 0 &0&2C_{33}\ema 
\label{eq:moduli}
\ee
superscripts $E$ recalling that strains are purely elastic.
The incremental response for various load directions is then computed, for different stress increments. We choose $\delta\vec\sigma$ values 
on a sphere in 3-space, centered at the origin, of radius $2\sqrt{2}\times 10^{-3} P$. Such increments are applied, and then multiplied by integer
factors 2, 3... up to 12, in order to record the influence of both their direction and their amplitude. The calculations are fully
stress-controlled, with variations of all 3 strain components. Once a new, pertubed equilibrium is reached, 
$\delta\vec\epsilon$ is measured, from which the elastic part $\delta\vec\epsilon^{E}=
\ww{C}^{-1}\cdot\delta\vec\sigma$ is subtracted, defining the 
\emph{irreversible} strain increment, which we denote with superscript $P$ (for ``plastic'').
Investigation points with $\sigma_2/\sigma_1=$ 1.2, 1.4, 1.6 and 1.8 were studied

Before presenting the results in the next section, let us recall that as a consequence of the assembling process, the investigated states possess a large 
coordination number, and (see Fig.~\ref{fig:biax}) are within the range of strain (``regime I''), along the biaxial loading curve,
that is dominated by contact deformation~\cite{Roux05,iviso3}. 
This means that the contact network does not break apart, and that the irreversible strains are due to sliding
at contacts where the Coulomb limit is reached. As a consequence, 
macroscopic strains, on changing confining stresses or stiffness constants, scale as $\kappa^{-1}$~\cite{Roux05}.
For larger deviators, or in poorly coordinated samples (which might be very dense nevertheless~\cite{iviso3}), the macroscopic
strains stem from network ruptures and rearrangements (``regime II'') and the incremental behavior might differ.
\section{Incremental response}
\subsection{No rotation of principal axes}
\begin{figure} 
\psfragscanon 
\psfrag{XTAG}{$\delta\sigma_1 / P$ }
\psfrag{YTAG}{$\delta\sigma_2 / P$ }
\includegraphics[width=0.36\textwidth]{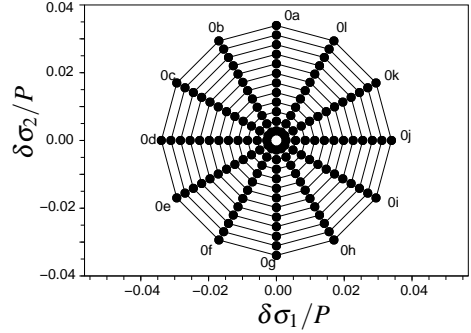}
\caption{Stress increments in principal axis plane} 
\label{fig:ChargeLevelPlots}
\end{figure}

\begin{figure} 
\psfragscanon 
\psfrag{XTAG}{ $\kappa\delta\epsilon_1^{E}$, $\kappa\delta\epsilon_1^{P}$ }
\psfrag{YTAG}{ $\kappa\delta\epsilon_2^{E}$, $\kappa\delta\epsilon_2^{P}$ }
\includegraphics[width=0.37\textwidth]{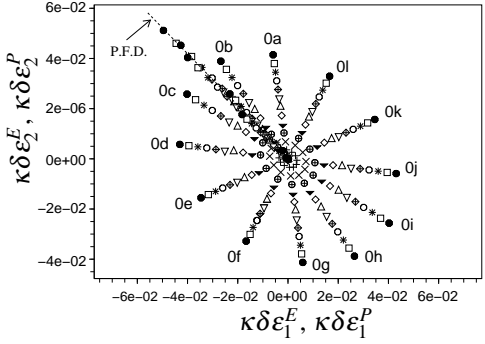}
\caption{Elastic and anelastic parts of response to stress increments marked (0a, 0b, ..., 0l) in
Fig.~\ref{fig:ChargeLevelPlots}}.
\label{fig:elastPlastPlot}
\end{figure}

We first investigate the response to stress
increments lying in the plane of principal stresses (i.e.,
$\delta\sigma_{3}=0$). For each investigation point along the curve of Fig.~\ref{fig:biax},
12 different orientations of $\delta\vec\sigma$ in this plane are tested, as shown in
Fig.~\ref{fig:ChargeLevelPlots}, with 12 different amplitudes (as specified before).
In order to
assess the relevance of classical plasticity models for the material
studied here we focus on the following three aspects: (i) the
existence of a flow rule dictating the direction of $\delta\vec\epsilon^{P}$; 
(ii) at equal amplitude $\vert\delta\vec\sigma\vert$, the linear dependence
of amplitude $\vert\delta\vec\epsilon^{P}\vert$ on the positive part $[\delta\hat\sigma]_+$ of 
$\delta\hat\sigma={\bf N}_C\cdot\delta\vec\sigma$, where ${\bf N}_C$ is the outer normal to some
yield criterion in stress space; (iii) the same linear dependence for varying stress
increment amplitudes.
The existence of a plastic flow rule is a sharp feature arising from
incremental tests, as shown in Fig.~\ref{fig:elastPlastPlot}, corresponding to
an investigation point with $\sigma_2 /\sigma_1 = 1.4$. 
Elastic strain increments $\delta\epsilon_1^{E}$ and $\delta\epsilon_2^{E}$
are disposed  along as many directions as the stress increments in
Fig.~\ref{fig:ChargeLevelPlots}, while plastic strain increments
($\delta\epsilon_1^{P}$ and $\delta\epsilon_2^{P}$) clearly align along a
unique direction, consistently with the flow rule. The same features are
observed for all investigation points. 

\begin{figure} 
\psfragscanon 
\psfrag{XTAG}{$\alpha$ }
\psfrag{XTAGSMALL}{\footnotesize$\alpha$ }
\psfrag{YTAG}{$\kappa\, \delta\epsilon^P$ }
\psfrag{S1AXIS}{\footnotesize$\delta\sigma_1$ }
\psfrag{S2AXIS}{\footnotesize\hspace{0.5mm}$\delta\sigma_2$ }
\psfrag{LAB1}{\footnotesize\hspace{1mm}$\alpha_{LD}$ }
\psfrag{LAB2}{\footnotesize\hspace{1mm}$\alpha_{PFD}$ }
\psfrag{LAB3}{\footnotesize\hspace{1mm}$\alpha_{NYC}$ }
\includegraphics[width=0.37\textwidth]{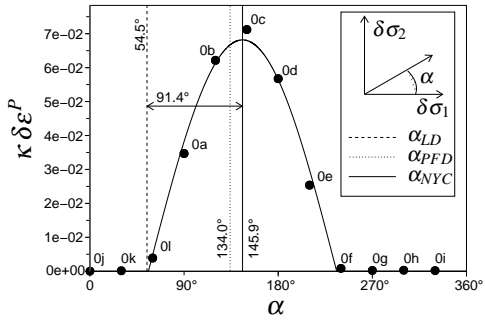}
\caption{Amplitude $\vert\delta\epsilon^P\vert$ vs. orientation $\alpha$ of 
  $\delta\vec\sigma$ for constant amplitude $\vert\delta\sigma\vert = 3.394 \cdot 10^{-2}$ ($\sigma_2/\sigma_1=1.4$).
  }
\label{fig:normYieldC}
\end{figure}

We discuss point (ii) of our list by referring to
Fig.~\ref{fig:normYieldC} in which $\vert\delta\vec\epsilon^P\vert$ is plotted
versus the angle $\alpha$ between principal axis 1 and increment $\delta\vec\sigma$, at
constant amplitude $\vert\delta\vec\sigma\vert$. In the framework of classical plasticity these values
should fit to the positive part of a cosine function reaching its maximum
in tnormal to the yield criterion.
Fitting theoretical curves to data allows to estimate the angle
$\alpha_{NYC}$ characterising the normal ${\bf N}_C$ to the yield criterion~\cite{TCV05} 
and the maximal amplitude $\delta\epsilon_{MAX}^P$ of the plastic strain increment.  
Notably, for all investigation points, the normal ${\bf N}_C$, oriented at angle 
$\alpha_{NYC}$, is consistently very nearly orthogonal to the current stress direction $\sigma_1, \sigma_2$ (oriented at angle $\alpha_{LD}$ 
in stress space). This suggests that the yield criterion might be defined by the Coulomb condition of a constant ratio $\sigma_2/\sigma_1$. 
Since  $\alpha_{PFD}\ne\alpha_{NYC}$ the plastic flow
direction differs from the normal ${\bf N}_C$, as in \emph{nonassociated elastoplasticity}.
\begin{figure}[!htb]
\psfragscanon
\psfrag{XTAG}{$[\delta\hat\sigma]_+$ }
\psfrag{YTAG}{$\kappa\, \vert\delta\vec\epsilon^P\vert$ }
\psfrag{TAG1}{\footnotesize$\sigma_2/\sigma_1=1.2$ }
\psfrag{TAG2}{\footnotesize$\sigma_2/\sigma_1=1.4$ }
\psfrag{TAG3}{\footnotesize$\sigma_2/\sigma_1=1.6$ }
\psfrag{TAG4}{\footnotesize$\sigma_2/\sigma_1=1$}
\includegraphics[width=0.36\textwidth]{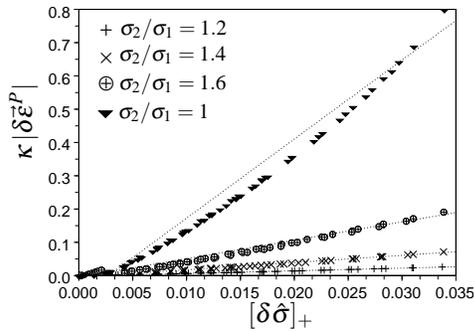}
\caption{Nonelastic strain amplitude vs. $[\delta\hat\sigma]_+$ defined with normal to
criterion identified in Fig.~\ref{fig:normYieldC}.
\label{fig:totalPlot}}
\end{figure}
As to point (iii), it is checked in Fig.~\ref{fig:totalPlot}, from which the following 
\emph{plastic moduli} $C_P$ (in units of $K_N$) are measured: $C_P= ??$, ??, ??, ?? corresponding, respectively, to
$\sigma_2/\sigma_1=$1.2, 1.4, 1.6 and 1.8. 
\subsection{General case}
\begin{figure}[!htb] 
\includegraphics[angle=270,width=5.5cm]{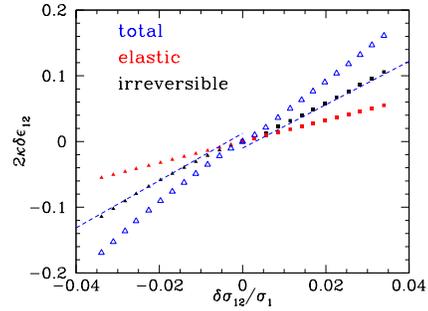}
\caption{\label{fig:ecroui33} Total, elastic, nonelastic shear strains as functions of applied shear stress to state with $\sigma_2/\sigma_1=1.8$.
Plastic modulus is close to $3K_N$ (resp. $2.8K_N$) for $\delta\sigma_{12}>0$ ($\delta\sigma_{12}<0$).} 
\end{figure}

\begin{figure}[!htb] 
\includegraphics[width=5.5cm]{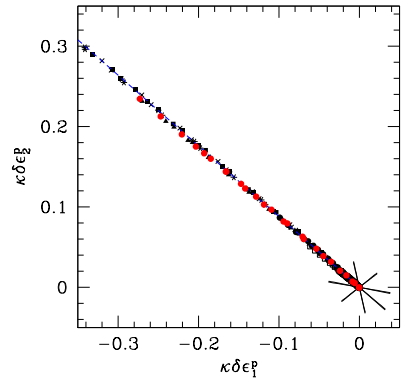}
\caption{\label{fig:dilatance2} Analog of Fig.~\ref{fig:elastPlastPlot} in state with $\sigma_2/\sigma_1=1.8$, 
for out-of-plane $\delta\vec\sigma$. Big red dots correspond to $\delta\sigma_1=\delta\sigma_2=0$. Elastic
strains (bottom right) are comparatively smaller.} 
\end{figure}

\begin{figure}[!htb] 
\includegraphics[width=5.5cm]{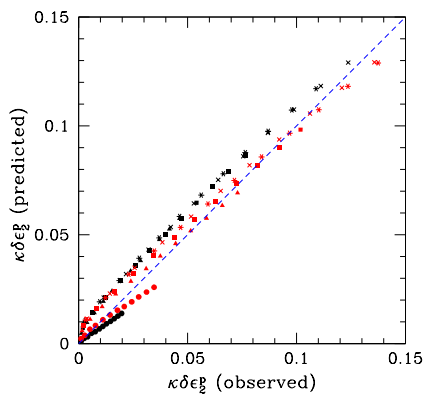}
\caption{\label{fig:combine} Predicted, with procedure defined in text, versus observed $\delta\epsilon_2$ for combined loads ($\sigma_2/\sigma_1=1.8$).} 
\end{figure}

If elastoplasticity applies -- which seems to be the case for $\delta\vec\sigma$ in the plane of the principal stress directions -- 
then a small load increment in the third direction, $\delta\sigma_3\ne 0$, $\delta\sigma_1=\delta\sigma_2=0$ should entail a purely elastic response.
Fig.~\ref{fig:ecroui33} contradicts this prediction, as a nonelastic shear strain $\delta\epsilon_3^{P}$ immediately appears, which increase proportionnally to 
shear stress $\vert\sigma_{12}\vert$. Coefficients can be slightly different for positive and negative $\delta\sigma_{12}$ because of finite sample size effects.
Like in-plane increments, such $\delta\vec\sigma$, if extremely small, yield a nonelastic response that is slightly sublinear in their amplitude, but a 
plastic modulus can be identified for $\delta\sigma_3/P$ of order $10^{-2}$.
Out-of-plane increments $\delta\vec\sigma$ also entail plastic strains $\delta\epsilon_1^{P}$, $\delta\epsilon_2^{P}$, which are still related by the same
flow rule as previously identified for in-plane loads ($\delta\sigma_3=0$). Fig.~\ref{fig:dilatance2} gathers results both from 16 load directions for which 
$\delta\sigma_3 = \pm \sqrt{\delta\sigma_1^2 + \delta\sigma_2^2}$, 
as well as simple shear increments ($\delta\sigma_1=\delta\sigma_2=0$) with both signs of $\delta\sigma_3$.
Quite surprisingly, the latter also produce a nonelastic reponse in the plane of principal stresses.
We thus observe that both the irreversible strains and the stress increments
causing them span two-dimensional spaces, with one in-plane and one out-of-plane direction, 
and that the response couples both directions. To be complete, we should then specify
how $\delta\vec\epsilon^P$ depends on $\delta\vec\sigma$ for all load increments. 
Although we are still investigating this issue, some preliminary attempts at superposition of
responses to shear and to in-plane stress increments are encouraging, as shown by Fig.~\ref{fig:combine}. 
Upon superimposing the previously identified responses to (in-plane) $\delta\hat\sigma={\bf N}_C\cdot\delta\vec\sigma$ 
and to $\vert\delta\sigma_3\vert$ in simple shear, 
Fig.~\ref{fig:combine} shows that the predicted values are fairly close to the measured ones.
\section{Microscopic aspects}
\begin{figure}[!htp]
\includegraphics[width=3.7cm]{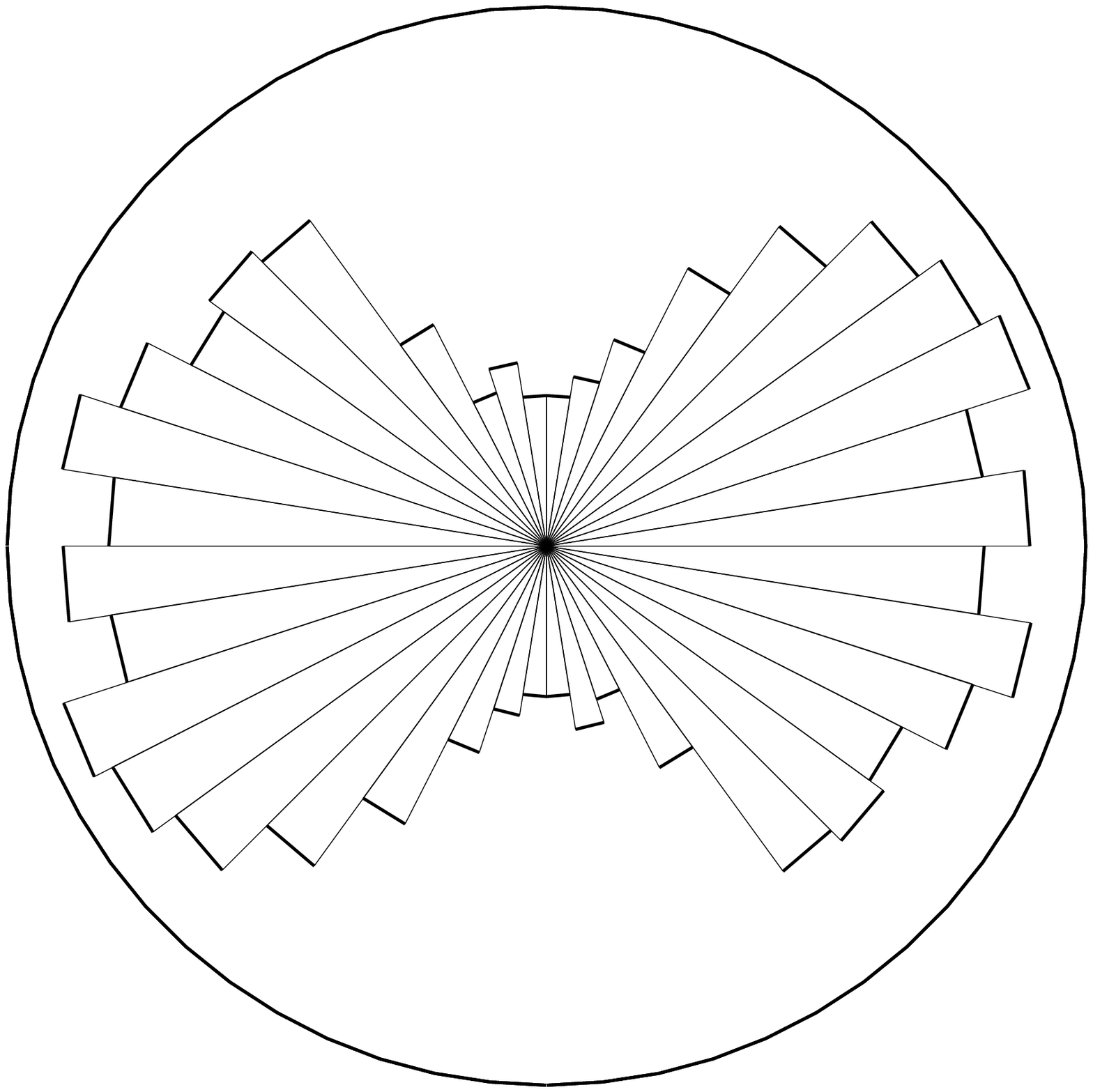}
\includegraphics[width=3.7cm]{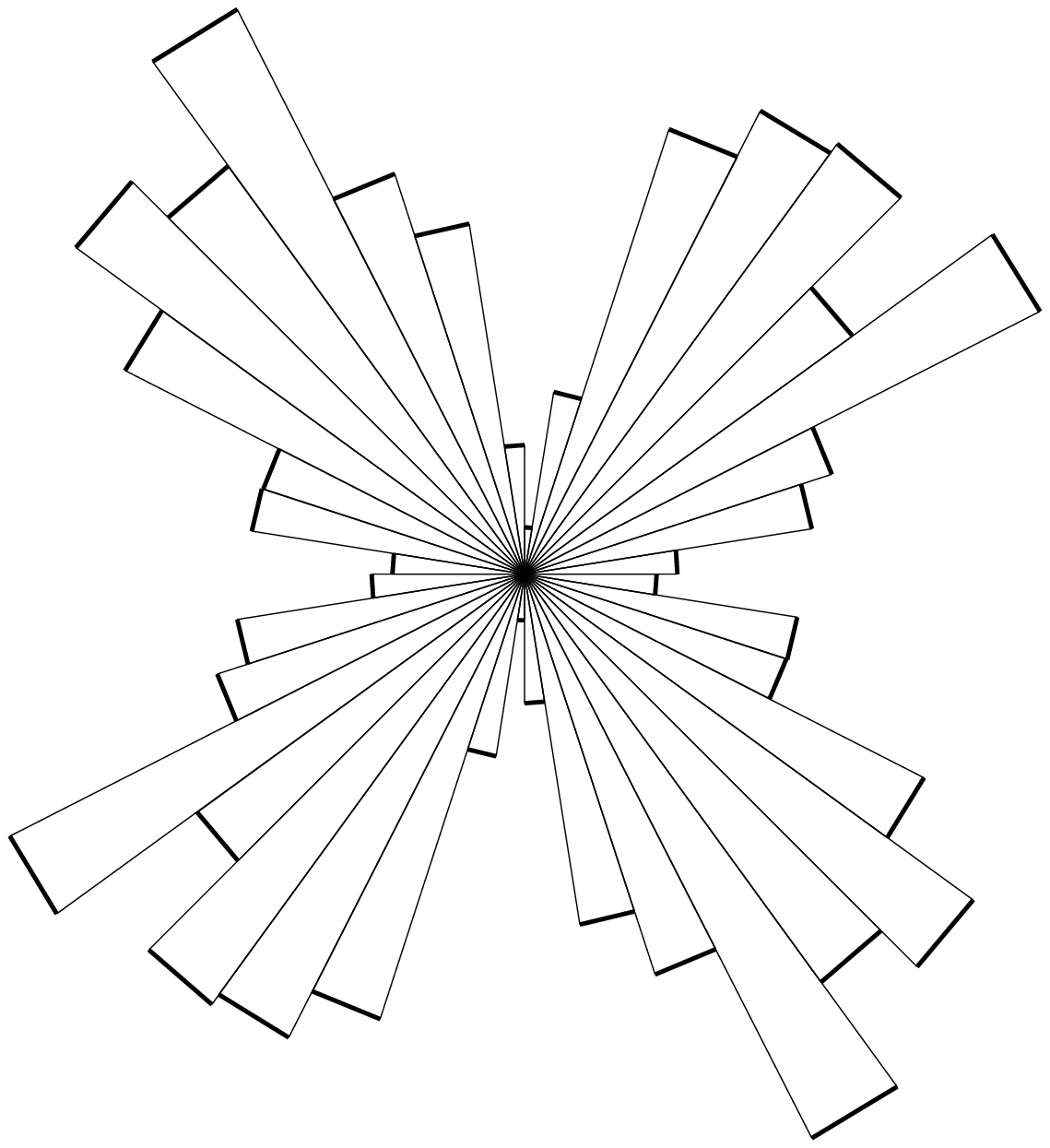}
\caption{\label{fig:fsliding}
Left: sliding contact orientational distribution (major principal axis vertical on the plot), 
normalized such that its angular average is a coordination number. Diameter of circle is 1.1,
global coordination is 3.5. Right: angular distribution of amplitude of sliding relative displacement in contacts in response
to in-plane load increment, normalized by plastic strain. $\sigma_2/\sigma_1=1.8$.
}
\end{figure}

The macroscopic nonelastic is due to plastic sliding in some contacts.
While the distribution of
contact orientations (fabric) is still moderately anisotropic in the investigated states, 
the sliding contact fabric (Fig.~\ref{fig:fsliding}) has a much stronger angular dependence. 
Such a distribution is observed for $\delta\hat\sigma>0$, while the population of sliding contacts virtually vanishes on applying
$\delta\hat\sigma<0$ and $\delta\sigma_3=0$. The sliding contact fabric depends on both $\delta\sigma_3$ and $\delta\hat\sigma$ in general. A nonzero
 $\delta\sigma_3$ breaks its symmetry. The angular distribution of sliding displacements at contacts (Fig.~\ref{fig:fsliding}), albeit different, is
also strongly anisotropic and shows similar sensitivity to the direction of $\delta\vec\sigma$. Finally, stress increments for which
$\delta\vec\sigma$ is proportional to $\vec\sigma$ (the neutral direction), entail no sliding, as
contact forces tend to increase proportionnally to their previous value.
\section{Perspectives}
The essential finding of the present study, which still remains to be systematized and calls for more thorough micromechanical investigations, is the
correspondence between 2D stress increments orthogonal to the currrent stress level and nonelastic strains belonging to a 2D space. In the near future
we plan to formulate it as a complete constitutive incremental law, to relate it to microscopic phenomena and to use it in localization criteria. 
The incremental response in systems with gradually rearranging contact networks (``regime II'', associated with microscopic instabilities) 
should also be investigated.
\bibliographystyle{aipproc}   


\begin{thebibliography}{11}
\expandafter\ifx\csname natexlab\endcsname\relax\def\natexlab#1{#1}\fi
\providecommand{\enquote}[1]{``#1''}
\expandafter\ifx\csname url\endcsname\relax
  \def\url#1{\texttt{#1}}\fi
\expandafter\ifx\csname urlprefix\endcsname\relax\def\urlprefix{URL }\fi
\providecommand{\eprint}[2][]{\url{#2}}

\bibitem[Vermeer(1998)]{VER98}
P.~A. Vermeer, in \emph{Physics of Dry Granular Media}, edited by H.~J. Herrmann,
  J.-P. Hovi, and S.~Luding, Balkema, Dordrecht, 1998, pp. 163--196.

\bibitem[Mitchell(1993)]{MIT93}
J.~K. Mitchell, \emph{Fundamentals of soil behavior}, Wiley, New York, 1993.

\bibitem[Alonso-Marroqu{\'\i}n et~al.(2005)]{AMLHV05}
F.~Alonso-Marroqu{\'\i}n, S.~Luding, H.~J. Herrmann, and I.~Vardoulakis,
  \emph{Phys. Rev. E} \textbf{71}, 051304 (2005).

\bibitem[Tamagnini et~al.(2005)]{TCV05}
C.~Tamagnini, F.~Calvetti, and G.~Viggiani, \emph{J. Eng. Math.} \textbf{52},
  265--291 (2005).

\bibitem[Darve et~al.({2007})]{DSDN07}
F.~Darve, L.~Sibille, A.~Daouadji, and F.~Nicot, \emph{{C. R. M\'ecanique}}
  \textbf{{335}}, {496--515} ({2007}).

\bibitem[Radja\"i(2008)]{R08}
F.~Radja\"i, \emph{ArXiv e-prints}  (2008), \eprint{0801.4722}.

\bibitem[Vardoulakis and Sulem(1995)]{VaSu95}
I.~Vardoulakis, and J.~Sulem, \emph{Bifurcation {A}nalysis in {G}eomechanics},
  Blackie {A}cademic and {P}rofessional, 1995.

\bibitem[Desrues and Chambon(2002)]{DeCh02}
J.~Desrues, and R.~Chambon, \emph{Int. J. Solid Struct.} \textbf{39},
  3757--3776 (2002).

\bibitem[Allen and Tildesley(1987)]{AT87}
M.~Allen, and D.~Tildesley, \emph{Computer simulations of liquids}, Oxford
  University Press, Oxford, 1987.

\bibitem[Peyneau and Roux(2008)]{PR08a}
P.-E. Peyneau, and J.-N. Roux, \emph{Phys. Rev. E} \textbf{78}, 011307 (2008).

\bibitem[Agnolin and Roux(2007)]{iviso3}
I.~Agnolin, and J.-N. Roux, \emph{Phys. Rev. E} \textbf{76}, 061304 (2007).

\bibitem[Roux(2005)]{Roux05}
J.-N. Roux,  in \emph{Powders and Grains 2005}, edited by R.~Garc\'{\i}a~Rojo,
  H.~J. Herrmann, and S.~McNamara, Balkema, Leiden, 2005, pp. 261--265.

\end{thebibliography}

\end{document}